\documentclass[aps,prl,twocolumn,groupedaddress,graphics,showpacs]{revtex4}
\usepackage{graphics}
\usepackage{epsfig}
\begin{document}
\bibliographystyle{apsrev}

\title{Tomonaga-Luttinger Liquid Features in Ballistic Single-Walled Carbon Nanotubes:
Conductance and Shot Noise}
\author{Na Young Kim}
\thanks{}
\email[]{Email: nayoung@stanford.edu}
\author{Patrik Recher}
\thanks{also at Institute of Industrial Science,
University of Tokyo, 4-6-1 Komaba, Meguro-ku, Tokyo 153-8505,
Japan}
\author{William D. Oliver}
\altaffiliation{Present address: MIT Lincoln Laboratory,
Lexington, Massachusetts, 02420}
\author{Yoshihisa Yamamoto}
\thanks{also at National Institute of Informatics, 2-1-2 Hitotsubashi, Chiyoda-ku,
Tokyo 101-8430, Japan} \affiliation{Quantum Entanglement Project,
SORST, JST, E. L. Ginzton Laboratory, Stanford University,
Stanford, California 94305}
\author{Jing Kong}
\altaffiliation{Present address: Department of Electrical
Engineering, MIT, Massachusetts, 02420}
\author{Hongjie Dai}
\affiliation{Department of Chemistry, Stanford University,
Stanford, California 94305 }
\date{October 5, 2006}

\begin{abstract}
We study the electrical transport properties of well-contacted
ballistic single-walled carbon nanotubes in a three-terminal
configuration at low temperatures. We observe signatures of strong
electron-electron interactions: the conductance exhibits
bias-voltage-dependent amplitudes of quantum interference
oscillation, and both the current noise and Fano factor manifest
bias-voltage-dependent power-law scalings. We analyze our data
within the Tomonaga-Luttinger liquid model using the
non-equilibrium Keldysh formalism and find qualitative and
quantitative agreement between experiment and theory.
\end{abstract}


\maketitle

Single-walled carbon nanotubes (SWNTs) continue to provide
numerous experimental and theoretical opportunities to investigate
one-dimensional physics upon their unique chemical, mechanical,
optical and electronic properties \cite{Dekker99}. Electrical
transport measurements with SWNTs have probed remarkable
electronic properties primarily via conductance. The ideal
conductance of SWNTs is $2G_{\text{Q}}= 2(2e^2/h)$ due to spin and
orbital degeneracy in principle, where $e$ is the elementary
charge and $h$ is Planck's constant; however, the measured
conductance is influenced by the quality of the contacts between a
tube and electrodes. For SWNTs weakly coupled to their electron
reservoirs (the tunnelling regime), the conductance exhibits a
power-law dependence on the drain-source voltage and/or
temperature as an indication of a Tomonaga-Luttinger liquid (TLL)
\cite{Bockrath99,Yao99}. In contrast, Pe{\c{c}}a \emph{et
al.}~theoretically analyzed SWNTs strongly coupled to the electron
reservoirs (the ohmic regime), claiming that the differential
conductance versus the drain-source bias voltage and the gate
voltage would unveil traits of the spin-charge separation
\cite{Peca03}. Recent theoretical efforts have sought to extend
the TLL analysis of ohmic SWNTs to include their current noise
properties \cite{Trauzettel0405, Lebedev05}. Corresponding
experimental observations in this regime have remained elusive.

Shot noise, non-equilibrium current fluctuations, originates from
the stochastic transport of quantized charged carriers. It probes
the second-order temporal correlation of electron current, which
often manifests certain microscopic physical mechanisms of the
conduction process. When Poisson statistics governs the emission
of electrons from a reservoir electrode, the spectral density of
the current fluctuations reaches its full shot noise spectral
density, $S=2eI$, where $I$ is the average current. In a
mesoscopic conductor, non-equilibrium shot noise occurs due to the
random partitioning of electrons by a scatterer, and it may be
further modified as a consequence of the quantum statistics and
interactions amongst charged carriers \cite{Blanter00}. A
conventional measure characterizing the shot noise level in
mesoscopic conductors is the Fano factor $F \equiv S/2eI$, the
ratio of the measured noise power spectral density $S$ to the full
shot noise value. Despite growing interest in the shot noise
properties of TLLs, current noise measurements in nanotubes have
only recently been executed due to the difficulty to achieve
highly-transparent (ohmic) contacts and a high signal-to-noise
ratio between the weak excess-noise signal and the prevalent
background noise \cite{Roche02}, although the shot noise
properties of SWNTs in the tunnelling regime with no TLL features
have been reported \cite{Onac06}.

In this letter, we address an experimental and theoretical study
of differential conductance and low-frequency shot noise with
well-contacted individual SWNTs at liquid ${}^4$He temperatures.
Experimental results on the differential conductance and
low-frequency shot noise reveal clear features of
electron-electron interaction. Quantum interference oscillation
amplitudes in differential conductance are strongly suppressed at
high bias voltages. In addition, the shot noise and the Fano
factor exhibit particular power-law scalings with the bias
voltage.

\begin{figure}
\epsfig{figure=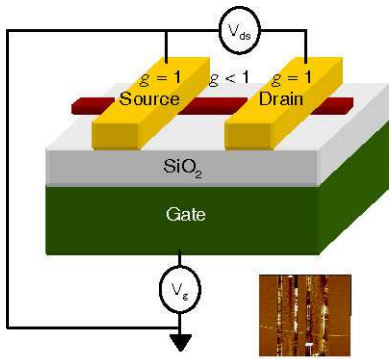,width = 2in} \caption{(Color online)
Illustration of a three-terminal SWNT device. The values of the
interaction parameter $g$ are indicated in the metal electrodes
($g = 1$) and in the SWNT ($g < 1$). Inset: atomic force
microscope image of a device. \label{Fig:fig1}}
\end{figure}

\begin{figure}
\epsfig{figure=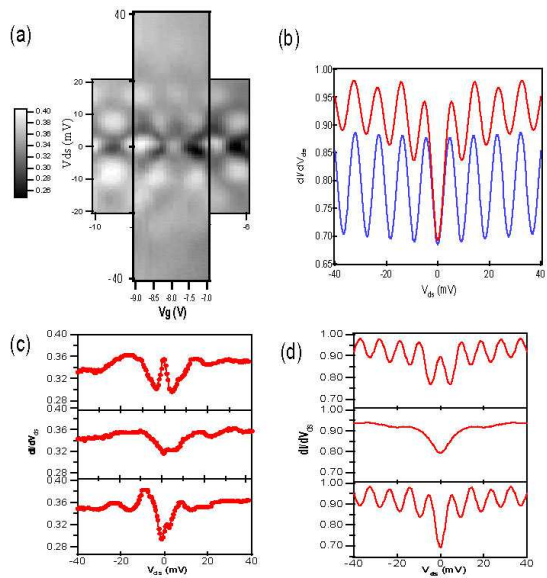,width = 2.8in} \caption{(Color online) The
differential conductance ($dI/dV_{\text{ds}}$) graphs in units of
2$G_{\text{Q}}$. (a) Density plot in the drain-source voltage
($V_{\text{ds}}$) and the gate voltage ($V_{\text{g}}$). 
(b) Theoretical $dI/dV_{\text{ds}}$ in $V_{\text{ds}}$ at a given
$V_{\text{g}}$ for $g$ = 1 (no interaction, blue) and $g$ = 0.25
(strong interaction, red) with $U_1 = 0.14$ and $U_2 = 0.1$ at $T$
= 4 K. (c) Three experimental traces at $V_{\text{g}}$ = - 9 V
(top), $V_{\text{g}}$= - 8.3 V (middle), and $V_{\text{g}}$ = -
7.7 V (bottom). (d) Theoretical traces at $T$ = 4 K for $U_2$ = -
0.1 (top), $U_2$ = 0 (middle) and $U_2$ = 0.1 (bottom) with
$U_1=0.14$. \label{Fig:fig2}}
\end{figure}

SWNT devices have a three-terminal geometry: source, drain, and
backgate as illustrated in Fig.~\ref{Fig:fig1}. The SWNTs were
synthesized using an Iron-based Alumina-supported catalyst with a
chemical vapor deposition method on a heavily doped Si substrate
with a 0.5 $\mu$m-thick thermal oxide \cite{Soh99}. The Si
substrate was used as the backgate. The metal electrodes were
patterned by electron beam lithography, defining the device
length. Ti/Au, Ti-only, and Pd metal electrodes were used, which
featured low-resistance contacts. Atomic force microscopy imaging
enabled us to select devices consisting of a single isolated SWNT
with 1.5 $\sim$ 3.5 nm diameter and 200 $\sim$ 600 nm length. We
measured the current - gate voltage ($V_{\text{g}}$) relation at
room temperature in order to distinguish metallic from
semiconducting tubes. The resistance of selected metallic SWNT
devices was typically 12 $\sim$ 50 k$\Omega$ at room temperature
and about 9 $\sim$ 25 k$\Omega$ at 4 K. SWNTs well-contacted to
two electrodes with finite reflection coefficients produced a
Fabry-Perot (FP) oscillation pattern in differential conductance
$dI/dV_{\text{ds}}$, ($V_{\text{ds}}$ is the drain-source voltage)
as an evidence of ballistic transport \cite{Liang01,Kong01}. Our
devices showed the FP interference at low $V_{\text{ds}}$, whose
diamond structures are caused by the confinement along the
longitudinal direction due to the potential barriers at the
interfaces with two metal electrodes (Fig.~\ref{Fig:fig2}(a)).
Contrary to the usual FP oscillations, we found in all devices
that the interference pattern fringe contrast reduced in magnitude
at high $V_{\text{ds}}$ as shown in Fig.~\ref{Fig:fig2}(a). This
feature cannot be explained by the standard Fermi liquid (FL)
theory, which predicts a constant oscillation amplitude regardless
of the bias voltage \cite{Liang01}.

Following Ref.~\cite{Peca03}, we model our device as a TLL with
two barriers separating the metal reservoirs from the SWNT and
with spatially inhomogeneous interaction parameter $g$. The
interaction is assumed to be strong in the SWNT $(0 < g < 1)$ and
weak in the higher dimensional metal reservoirs ($g = 1$). The TLL
without the barriers is described by the bosonized Hamiltonian
\cite{Kane97} $H_{\text{\textsc{swnt}}} =
(v_{\text{F}}/2\pi)\sum_{a}\int
dx[(\partial_{x}\phi_{a})^2+g_{a}^{-2}(x)(\partial_{x}\theta_{a})^2]$,
where $\theta_{a}(x)$ and $\Pi_{a}(x) = -\partial_{x}\phi_{a}/\pi$
are conjugated bosonic variables, i.e.
$[\theta_{a}(x),\Pi_{b}(x^\prime)] = i
\delta_{ab}\delta(x-x^\prime)$, and $v_{\text{F}}$ is the Fermi
velocity. The four conducting transverse modes of the SWNT in the
FL theory are transformed to four collective excitations  in the
TLL theory: one interacting collective mode ($a=1, g_{a} \equiv
g$) of the total charge and three non-interacting collective modes
($a=2-4, g_{a} = 1$) including spin. These modes are partially
reflected at the two barriers. The interacting mode further
experiences momentum-conserving reflections due to the mismatch of
$g$ at the interfaces \cite{Safi95}. We compute
$dI/dV_{\text{ds}}$ using the Keldysh formalism and treat the
barriers as a weak perturbation. We obtain $I \equiv
e(2/\pi)\dot{\theta_{1}}= 2G_{\text{Q}}V_{\text{ds}}-I_{\text{B}}$
where $I_{\text{B}}$ is given to leading order in the
backscattering amplitudes as \cite{Peca03,Recher06}
\begin{equation}
\label{equation2} I_{\text{B}} = \frac{2e}{\pi t_{\text{F}}^2}
\sum_{n = 1,2} U_{\text{n}} \left|\int_{0}^{\infty} dt
e^{\text{C}_{\text{n}}(t)}
\sin\left(\frac{\text{R}_{\text{n}}(t)}{2}\right) \sin\left(\frac{
eV_{\text{ds}}t}{\hbar}\right)\right|,
\end{equation}
where $t_{\text{F}} = L/v_{\text{F}}$ is the travelling time for a
non-interacting mode along the SWNT length $L$. The backscattered
current $I_{\text{B}}$ consists of two contributions: the term
proportional to $U_1$ represents the incoherent sum of
backscattering events at the two barriers and the term associated
with $U_{2}$ results in the FP oscillations due to the coherent
interference between backscattering events from different
barriers. At high $V_{\text{ds}}$, the $U_1$-term in
Eq.~(\ref{equation2}) dominates and the oscillation amplitude
decreases. $U_{1}$ and $U_{2}$ are independent of $V_{\text{ds}}$,
but $U_2$ depends periodically on $V_{\text{g}}$
\cite{amplitudenote}. The interaction parameter $g$ is involved in
the time integral through $\text{C}_{\text{n}}(t)$ and
$\text{R}_{\text{n}}(t)$, which are correlation and retarded
functions of the fields $\theta_a$ and $\phi_a$, respectively.
These Green's functions contain a sum
over all four collective modes 
and their forms are obtained at zero \cite{Peca03, Recher06} and
finite temperatures \cite{Recher06}.

Figure~\ref{Fig:fig2}(b) contrasts the effect of electron-electron
interaction ($g$ = 0.25 (red)) on $dI/dV_{\text{ds}}$ with its
non-interacting counterpart ($g$ = 1 (blue)) for a SWNT of length
$L$ $\sim$ 360 nm with fitting parameters $U_1 = 0.14$ and $U_2 =
0.1$. The amplitude of the FP oscillation is damped
at high $V_{\text{ds}}$ compared to that at low $V_{\text{ds}}$.
Experimental traces show the trend of the FP oscillation amplitude
reduction as predicted by the TLL theory for $g$ = 0.25; however,
the overall conductance of real devices was lower than that in
theory. To identify the TLL feature uniquely in experiments
requires one to increase $V_{\text{ds}}$ above the levelspacing
$\hbar/2gt_{\text{F}}$.  Note that the tendency of amplitude
reduction in experimental data cannot be reproduced by the
reservoir heating model \cite{Henny99a} which asserts that the
dissipated power $V_{\text{ds}}^2 (dI/dV_{\text{ds}}) $ leads to a
bias-voltage dependent electron temperature \cite{Liangfootnote}.
We have tested this effect for the non-interacting case ($g$ = 1)
in our theory and have found that it causes a slight damping of
the FP-oscillations ($U_2$-term) but the incoherent part
($U_1$-term) is independent of temperature \cite{Recher06}. The
temperature effect, therefore, fails to account for the
experimentally observed enhanced backscattering amplitude at low
$V_{\text{ds}}$. In addition, the conductance is relatively small
(on the order of $G_Q$) so that heating effects should not be
pronounced in the bias window considered.

Figure~\ref{Fig:fig2}(c) presents several traces in
$V_{\text{ds}}$ at different $V_{\text{g}}$, from which the
following pronounced features are observed: the period of the
oscillations at low $V_{\text{ds}}$ depends on the value of
$V_{\text{g}}$, and it becomes elongated at high $V_{\text{ds}}$.
The former feature can be explained by the TLL model. The model
states that the period is $2g t_{\text{F}}$ if $U_{2}\sim 0$ since
the non-zero contribution to the oscillation is only from the
interacting mode while the oscillations from the three
non-interacting modes destructively interfere
[Fig.~\ref{Fig:fig2}(d)(middle)]. The dominant period is
$t_{\text{F}}$ produced by the three non-interacting modes when
$U_{2}$ is maximal (e.g. $U_2 = \mp 0.1$ in
Fig.~\ref{Fig:fig2}(d)(top, bottom)). The $V_{\text{g}}$-dependent
oscillation periods in $dI/dV_{\text{ds}}$ have been interpreted
as a signature of spin-charge separation in the SWNT
\cite{Peca03}. We find an indication of this effect by comparing
the primary periods of these traces, yielding $g \sim 0.22$. The
latter feature observed, a longer period at high $V_{\text{ds}}$,
is beyond our theory, but it is likely to be caused by a strong
barrier asymmetry at high $V_{\text {ds}}$ which would also
suppress the $U_{2}$-term. The ratio of primary to elongated
periods along $V_{\text{ds}}$ gives $g \sim 0.22$ as well.
Although compelling evidence, further experiments focusing on the
periodicity with $V_{\text{ds}}$ should be performed to be
conclusive.

The shot noise measurements were performed by placing two current
noise sources in parallel: a SWNT device and a full shot noise
generator. The full shot noise standard is a weakly coupled light
emitting diode (LED) and photodiode (PD) pair. In our 4-K
implementation, the overall coupling efficiency from the LED input
current to the PD output current was about 0.1 $\%$, which
eliminated completely the shot noise squeezing effect due to
constant current operation \cite{Kim97}. In order to recover the
weak shot noise embedded in the background thermal noise, we
implemented an AC modulation lock-in technique and designed a
resonant tank-circuit together with a home-built cryogenic
low-noise preamplifier \cite{Reznikov95, Liu98, Oliver99}. The
input-referred voltage noise of the circuit was approximately 2.2
nV/$\sqrt{\text{Hz}}$ at 4 K with a resonant frequency $\sim$ 15
MHz. The preamplified signal was fed into a room-temperature
amplifier with a gain of about 30 dB, a bandpass filter with low
and high cutoff frequencies of 12 and 21.4 MHz, a square-law
detector, and a lock-in amplifier. The Fano factor $F(I_i) \equiv
S_{\text{\textsc{swnt}}} (I_i)/S_{\text{\text{\textsc{pd}}}}
(I_i)$ was obtained from the ratio of the SWNT current noise
spectral density ($S_{\text{\textsc{swnt}}}$) to the LED/PD full
shot noise spectral density ($S_{\text{\textsc{pd}}}$) at each dc
current value $I_i$. The current noise generated in the LED/PD
pair was measured while the SWNT was dc voltage-biased with a
constant dc current $I_i$.

\begin{figure}
\epsfig{figure=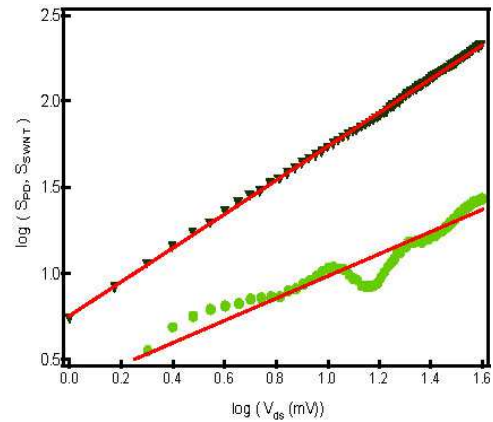,width=2.5in} \caption{(Color online) Shot
noise power spectral density versus $V_{\text{ds}}$ for the LED/PD
pair ($S_{\text{\textsc{pd}}}$, triangle) and the SWNT
($S_{\text{\textsc{swnt}}}$, dot) at $V_{\text{g}}$ = - 7.9 V. The
slopes of the shot noise spectral densities are 1 and 0.64 for
LED/PD and SWNT respectively, from which the inferred $g$ value
for SWNT is 0.16. \label{Fig:fig3}}
\end{figure}

Figure~\ref{Fig:fig3} shows a typical log-log plot (base 10) of
$S_{\text{\textsc{swnt}}}$ in $V_{\text{ds}}$ at a particular
$V_{\text{g}}$. $S_{\text{\textsc{swnt}}}$ (dot) is clearly
suppressed to values below full shot noise
$S_{\text{\textsc{pd}}}$ (triangle), and it
suggests that the
relevant backscattering for shot noise is indeed weak. Note that
$S_{\text{\textsc{swnt}}}$ and $S_{\text{\textsc{pd}}}$ have
clearly different scaling slopes versus $V_{\text{ds}}$.

We extend the theory to calculate the shot noise spectral density,
$S_{\text{\textsc{swnt}}}(\omega) = \int dt e^{i \omega t} \langle
\{ \delta \hat{I}(t), \delta \hat{I} (0)\} \rangle $ with $\delta
\hat{I}(t) = \hat{I}(t)-I$ the current fluctuation operator and
$\{\cdots\}$ the anticommutator \cite{Recher06}. The SWNT noise in
the zero-frequency limit is expressed as $S_{\text{\textsc{swnt}}}
= 2e\coth(eV_{\text{ds}}/2k_{\text{B}}T)I_{\text{B}}+
4k_{\text{B}}T
(dI/dV_{\text{ds}}-dI_{\text{B}}/{dV_{\text{ds}}})$, becoming
$S_{\text{\textsc{swnt}}} = 2eI_{\text{B}}$ for $eV_{\text{ds}} >
k_{\text{B}}T$. The asymptotic behavior of $I_{\text{B}}$ from the
dominant $U_1$-term when $eV_{\text{ds}}
> \hbar/2gt_{\text{F}}$ follows the power-law scaling $I_{\text{B}}
\sim V_{\text{ds}}^{1+\alpha}$ with $\alpha = -(1/2) (1-g)/(1+g)$.
Note that $\alpha$ is uniquely determined by the TLL parameter
$g$. The average value of the power exponent in this sample
(Fig.~\ref{Fig:fig3}) over seven different gate voltages is
estimated to be $\alpha \sim -  0.31 \pm 0.027$, corresponding to
$g \sim 0.25 \pm 0.049$.

\begin{figure}
\epsfig{figure=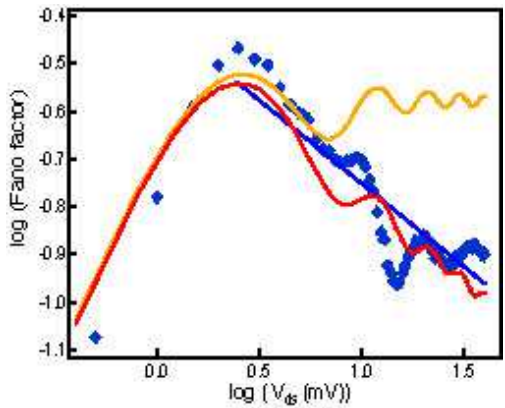,width=2.6in} \caption{ (Color online) Fano
factor versus $V_{\text{ds}}$ on a log-log scale. The theoretical
Fano factor curves where thermal noise
4$k_{\text{B}}T\left(dI/dV_{\text{ds}}\right)$ is subtracted are
drawn for $g$ = 1 (yellow) and $g$ = 0.25 (red) at T = 4 K with
$U_1 = 0.14$ and $U_2 = 0.1$. The power exponent $\alpha \sim$ -
0.35 (blue line) for the measured Fano factor (diamond) at
$V_{\text{g}}$  = - 7.9 V, and the inferred $g$ value is 0.18. The
theoretical $g$ = 1 (yellow) plot gives $\alpha \sim$ 0 as
expected. \label{Fig:fig4}}
\end{figure}

The experimental Fano factor $F(I)$ is displayed on a log-log
(base 10) scale in Fig.~\ref{Fig:fig4}. The TLL model predicts
that at low bias voltages $eV_{\text{ds}} < k_{\text{B}}T <
\hbar/2gt_{\text{F}}$, $F(I) \propto V_{\text{ds}}$ after
subtracting the thermal noise component
$4k_{\text{B}}T\left(dI/dV_{\text{ds}}\right)$, and its slope is
insensitive to $g$-values in the region of $\log(V_{\text{ds}}) <
\log(\hbar/2get_{\text{F}}) \sim 0.47$ for a 360 nm SWNT and $g =
0.25$ (Fig.~\ref{Fig:fig4}). On the other hand, if $eV_{\text{ds}}
> \hbar/2gt_{\text{F}}$, a power-law $F \sim V_{\text{ds}}^\alpha$
is expected by assumption that 
the backscattered current is smaller than the ideal current
2$G_{\text{Q}}V_{\text{ds}}$. 
A linear regression analysis of the Fano factor $F$ with
$V_{\text{ds}}$ in this region, therefore, is another means to
obtain the $g$ value. The Fano factors $F$ for $g$ = 0.25 (red)
and $g$ = 1 (yellow) are displayed on a log-log scale in
Fig.~\ref{Fig:fig4}. The experimental data (diamonds) agree well
with the theoretical Fano factor of $g$ = 0.25. The stiffer slope
($\alpha$) corresponds to stronger electron-electron interaction.
The mean value of the exponent $\alpha$ and the inferred $g$
derived over seven $V_{\text{g}}$ values are $\alpha = - 0.33 \pm
0.029$ and $g = 0.22 \pm 0.046 $ respectively for this particular
sample. We find that the measured exponents $\alpha$ and inferred
$g$ values from the spectral density $S_{\text{\textsc{swnt}}}$
and the Fano factor $F$ from four different devices with various
metal electrodes (Ti/Au, Ti-only, Pd) show similar statistics
$\alpha \sim - 0.31 \pm 0.047$ and $g \sim 0.26 \pm 0.071$ as
derived from several $V_{\text{g}}$ values for each sample. We
stress that the non-linear decay of the experimental $F$ along
$V_{\text{ds}}$ indeed starts at a voltage scale
$\log(\hbar/2get_{\text{F}})\sim 0.61$ for $g \sim 0.18$ in Fig. 4
as a manifestation of a collective electron effect.

We have measured non-equilibrium differential conductance and shot
noise in ballistic SWNTs at low temperatures and analyzed the data
within the TLL theory including weak electron backscattering at
the SWNT-metal reservoir interfaces. We find convincing agreement
between experiment and theory: reduced conductance
(FP)-oscillation amplitudes with increasing bias voltage, and
power-law characteristics in the weak backscattered current
component through low-frequency shot noise measurements. The joint
measurement of differential conductance and shot noise provides
independent experimental access to the transmitted and
backscattered current components of the non-equilibrium transport.
This measurement constitutes the first quantitative investigation
of TLL interaction effects in the shot noise of SWNTs.

 We acknowledge Prof. Quate for his support and atomic force
microscopes for imaging the SWNT devices, A. Javey for
Pd-contacted devices, and C. Sch\"onenberger, H. Grabert and B.
Trauzettel for helpful discussions and comments. This work was
supported by the ARO-MURI grant DAAD19-99-1-0215, JST/SORST, NTT,
and the University of Tokyo.


\begin{references}{}
\bibitem{amplitudenote}
We find $U_1\propto\sum_{ij}[(u_{1}^{ij})^2+(u_{2}^{ij})^2]$ and
$U_{2}\propto 2\sum_{ij}u_{1}^{ij}u_{2}^{ij}\cos(\eta
V_{g}+2\Delta_{ij})$ where $u_{1,2}^{ij}$ are the bare
backscattering amplitudes for barrier 1,2 and between subbands
$i,j$ of the SWNT, $\Delta_{ij}$ are the corresponding scattering
phases and $\eta$ is a device-dependent constant.
\bibitem{Liangfootnote}
Heating was proposed as the likely cause for the diminished
FP-amplitude with increasing $V_{\text{ds}}$ in
Ref.~\cite{Liang01}.
\end{references}
\end{document}